\documentclass{article}
\usepackage{ijcai20}

\usepackage{times}
\usepackage{soul}
\usepackage{enumitem}
\usepackage{url}
\usepackage[hidelinks]{hyperref}
\usepackage[utf8]{inputenc}
\usepackage[small]{caption}
\usepackage{graphicx}
\usepackage{amsmath}
\usepackage{amsthm}
\usepackage{booktabs}
\usepackage{algorithm}
\usepackage{algorithmic}
\title{Whither Fair Clustering?}

\author{
    Deepak P
    \affiliations
    Queen's University Belfast, UK
    \emails
    deepaksp@acm.org
}

\begin{document}

\maketitle

\begin{abstract}
Within the relatively busy area of fair machine learning that has been dominated by classification fairness research, fairness in clustering has started to see some recent attention. In this position paper, we assess the existing work in fair clustering and observe that there are several directions that are yet to be explored, and postulate that the state-of-the-art in fair clustering has been quite parochial in outlook. We posit that widening the normative principles to target for, characterizing shortfalls where the target cannot be achieved fully, and making use of knowledge of downstream processes can significantly widen the scope of research in fair clustering research. At a time when clustering and unsupervised learning are being increasingly used to make and influence decisions that matter significantly to human lives, we believe that widening the ambit of fair clustering is of immense significance. 
\end{abstract}

\section{Introduction}

Fair Machine Learning (Fair ML) is a flourishing discipline of study that has gathered much attention in the last several years, starting from an early pioneering work in~\cite{dwork2012fairness}. Of late, a newly instituted interdisciplinary conference series, ACM FAT*/FAccT\footnote{https://facctconference.org/} has bolstered further interest. Broadly, there have been two fairness streams explored in Fair ML literature: (i) {\it individual fairness} that prefers adherence to {\it treating similar people similarly}, and (ii) {\it group fairness} which involves ensuring some notion of {\it 'fair'} distribution of analytics results across groups defined on {\it sensitive attributes} such as gender, race, ethnicity and religion. Over the past years, significant progress has been made in fair classification, with emergence of computational notions such as {\it independence}, {\it separation} and {\it sufficiency}~\cite{barocas2017fairness}. Supervised learning has the luxury of availability of labelled data that encompasses information of historical decisions. In the case of binary decision making (success/fail), the chasm between the base success rate for each sensitive class (e.g., gender) and their representation within the training data provides a fertile ground for the pursuit of fairer supervised learning. On the other hand, unsupervised or exploratory learning does not assume availability of labels in the data, making fairness within unsupervised learning quite a distinct notion from the former. It may be noted that unsupervised learning is of growing significance in ML, and is often referred to as the {\it next frontier in AI}\footnote{https://bit.ly/2zWjTEo - Yann LeCun, 2018 Turing Laureate}.

Fairness in unsupervised machine learning may be expected to increase in importance with the broadening scope of unsupervised learning, facilitated by the growth of data volumes far outpacing any attempt at getting them labelled. This data growth has been facilitated in the public sector by an expansion of the methods for ’passive’ data collection, where data is collected through safety/surveilance cameras and IoT devices as part of smart city infrastructure. In the private sector, the user's mobility patterns are available to map services (e.g., Google Maps, Bing Maps) and black-box car insurance providers, social interests are available to social media companies (e.g., Facebook, Twitter), and with the advent of PDAs (e.g., Echo, Home), web tech giants can potentially have access to audio conversations within homes. 

%, various kinds of sensors, smartphone apps, medical wearables, traffic sensing devices, public wifi and even social media monitoring. 

%fashion interests to e-commerce websites (e.g., Amazon), physical activity data to fitness tracking providers (e.g., Fitbit) as well as to medical wearable clouds, dining interests to the likes of Yelp, FMCG/CPG purchasing activities to supermarkets (e.g., Tesco, Walmart) through loyalty programmes,

%This is to be contrasted with active data collection that is the basis of most administrative data, such as with government forms completed by citizens and public servants, which typically get some form of manual labelling at the point of submission. 

Clustering, arguably the most popular task in unsupervised learning, has seen much fairness-oriented research attention in the last few years. The pioneering work on this stream was one on data pre-processing to facilitate fair clustering~\cite{chierichetti2017fair}. Our analysis of the community's approach to the task across the 15+ papers in literature leads us to an argument that the literature has been quite restricted in scope. This is arguably due to treating it as a well-defined computational task, despite it being much more nuanced due to being situated within the space of a sophisticated landscape of normative principles. We assess fair clustering literature in the backdrop of the political philosophy around fairness and justice, and make the following arguments:

\begin{itemize}[leftmargin=*]
    \item {\bf Normative Target:} The normative principles, the space of values targeted, that have been used across fair clustering formulations have been quite narrow in scope, and significantly narrower than in the case of fair supervised learning. This is to be seen in the backdrop of the plethora of normative principles available in political philosophy. In particular, we observe that most clustering formulations have relied on {\it alleviating disparate impact through representational parity}, a pursuit of {\it group fairness} that relates to {\it luck egalitarianism}~\cite{lang2009luck} when sensitive attributes are considered as manifestations of {\it brute luck} choices. It is also noteworthy that the relationship between egalitarianism and discrimination avoidance has been argued to be nuanced~\cite{binns2018fairness}. 
    \item {\bf Shortfall Characterization:} Clustering, as a dataset-level optimization task, is very well understood to be complex. Given the complexities, most formulations fall short of achieving the representational parity goal that they target for. Techniques have focused on either bounding the shortfall theoretically, or illustrating empirically that the quantum of shortfall is tolerable. The critical missing piece is that the shortfall, while being quantified as above, has been left uncharacterized. It has not been elucidated as to what {\it what kind of data objects are likely to suffer more or less} from the shortfall. For usage in practical scenarios, especially within public sector, absence of such a characterization of the shortfall could be a potential dealbreaker. 
    \item {\bf Application Space:} Most clustering formulations seek to achieve their fairness goal in each of the clusters in the output. In a way, they are being application-agnostic and try to ensure that {\it whatever be the downstream application that makes use of the clustering, there is some form of fairness assurance} that the techniques provide. However, typical clustering outputs could be used in order to decide from among a small set of decision choices, which could additionally be placed somewhere in the spectrum of positive or negative. Information about the downstream usage of clustering outputs could both: (i) improve the ability to optimize better for the chosen optimization goal, and (ii) render the formulation more suited to particular domains. 
\end{itemize}

\section{Case Study: Clustering for Job Shortlisting}

Towards putting forth the arguments raised above, we will use the backdrop of a setting where clustering is used to inform consequential decisions directly. Consider the case of a heavily oversubscribed job vacancy, where manual perusal of each of them is out of question. Such a scenario is routinely encountered in the case of government jobs in populated developing countries\footnote{https://www.bbc.co.uk/news/world-asia-india-43551719}. We consider a pipeline of clustering usage for such a scenario. {\it First}, the received applications would be subject to clustering using a similarity measure that is relevant to assessing the suitability to the job, to generate perhaps hundreds of clusters. {\it Second}, a representative application from each cluster, perhaps the {\it medoid}, would be subject to manual assessment for suitability to the job. {\it Third}, the arrived assessment for the medoid, likely one of {\it shortlist, reject, scrutinize further} would be applied to all applications in its cluster. {\it Fourth}, those labelled {\it scrutinize further} by virtue of enough ambiguity on the suitability assessment, could be subject to further clustering, or if there is enough manual bandwidth available, subject to individual manual assessments. As an illustrative example to appreciate the need for clustering fairness within this pipeline, observe that generating a set of gender-skewed clusters could help reinforce gender stereotypes that play a part in manual perusal. The cluster-level decisions made over such gender-skewed clusters could then become implicitly gender-aligned. Data analytics' role in reinforcing social and economic inequalities has been the topic of several recent books~\cite{o2016weapons}. 

%In this scenario, it is easy to see how unfair clustering could deliver disparate impact. 

\section{Normative Target: What to Optimize for}

The normative principle used in a number of fair clustering formulations is that of assuming that each attribute be either considered {\it sensitive} or {task-relevant}, followed by {\it targeting to preserve the dataset-wide distribution of objects along sensitive attributes within each cluster}. For example, with gender regarded sensitive, this translates to ensuring that the gender ratio within each cluster be very similar to the gender ratio in the dataset. Different fair clustering formulations differ in the {\it number} and {\it kind} of sensitive attributes they admit; such a characterization of literature appears in~\cite{DBLP:conf/edbt/Abraham0S20} (Ref. Table 1). The similarities between data objects on task-relevant attributes are deemed to be relevant to the task {\it in the same manner}; weights may be attached to attributes to differentiate the quantum of influence, but the {\it nature} of the influence remains similar. %We enumerate illustrative directions in which fair clustering could target different normative formulations. 

{\it First}, the crisp binary distinction between sensitive and non-sensitive attributes begs apparent criticism. There are often attributes on which discrimination could be avoided, but not necessarily as strongly. For example, the {\it age} or {\it region/province} attribute could be such; there is typically a higher degree of tolerance towards skew in age and regions (e.g., urban skew), but purely age-homoegeous or region-homogeneous clusters are nevertheless undesirable. {\it Second}, while sensitive attributes are often outcomes of what are called {\it brute luck}, there exist other luck-influenced attributes whose placement is not clear in the sensitive/task-relevant dichotomy. These include the likes of {\it option luck}~\cite{dworkin2002sovereign} which relate to choices made on the face of considerable uncertainty of how things would turn out. For example, a career-break due to startup failure is unlike {\it brute luck}, but still not something that the candidate should heavily scored down on. Some addressal of option luck may be achieved by manually engineering covariate features to control. {\it Third}, there is significant space to expand the normative target outside the space of {\it egalitarianism}, notably the Rawlsian choice~\cite{rawls1971theory} in the fairness-efficiency trade-off. There are other targets within the so-called {\it patterned notions} (as outlined in~\cite{nozick1974anarchy}), and prefer to allocate resources in accordance with patterns such as {\it need} or {\it moral desert}\footnote{Desert (in philosophy) $\approx$ quality of being considered deserving.}. This would be especially true of hiring in the public sector where the government could use such patterned allocation in order to associate esteem with certain values. This would require identifying attributes that correlate with {\it need} and {\it desert} and treating them specially so that people with similar needs and deserts be clustered together. Desert may often need to be specified through attribute-combinations; a candidate from a backward region who has shown exceptional interest in a trade despite limited access to facilities may be considered as scoring high on moral desert. Similarity search has explored multiple unconventional and complex aggregation operators~\cite{deepak2015operators}. The lack of diversity in normative targets is also true of supervised machine learning, though perhaps only to a lesser extent. 

% For example, the extent of the moral desert may be a non-trivial function of multiple attributes and sub-spaces within them

While we started off observing that group fairness on sensitive attributes has been the mainstay in fair clustering, a few deviant formulations are worthy of mention. Proportionally fair clustering~\cite{pmlr-v97-chen19d} proposes an ingenious notion of {\it collective desert}; it requires that a sufficiently large collective of proximal objects would deserve a cluster of their own. {\it Representativity Fairness}~\cite{repfairness}, on the other hand, prefers egalitarian distribution of the {\it cost of abstraction} incurred due to the clustering.% equally across objects in the dataset. %is independent of sensitive attributes

\section{Characterization of Residual Unfairness}

Once the normative target is decided, fair clustering formulations translate the target to a mathematical optimization formulation. With even simple clustering formulations being computationally hard~\cite{mahajan2012planar}, fair clustering will also involve approximations. These might be in the form of theoretical approximation bounds~\cite{chierichetti2017fair,NIPS2019_8741} or demonstration of empirical effectiveness~\cite{DBLP:conf/edbt/Abraham0S20}. While it is eminently desirable that the chosen target be achieved as much as possible, it is also useful to have an understanding of {\it how} it falls short when it does indeed fall short; this aspect has not been explored at all to our best knowledge. An important question that one may ask is whether the residual unfairness is {\it Rawlsian}~\cite{rawls1971theory}; whether it is arranged to the greatest benefit of the least advantaged (ref. {\it difference principle}). Answers to such questions are crucial for uptake in practical applications since some kinds of systematic unfairness may be considered as intolerable, especially within public sector. 

%it unfavorable to mixed-size clusters in the data, with This makes objects in the fringes of the large clusters being highly susceptible to being {\it 'engulfed'} by the smaller clusters. 

Consider the immensely popular $K$-Means formulation for clustering~\cite{macqueen1967some}, which some fair clustering formulations build upon (e.g.,~\cite{DBLP:conf/edbt/Abraham0S20,ziko2019clustering}). $K$-Means clusters may be seen as being located within Voronoi cells centered on the cluster means. Since fair clustering algorithms building upon the $K$-Means framework are intuitively likely to make the {\it pro-fairness adjustments} through membership re-assignments at the fringes of clusters, fringe objects would likely bear the {\it cost/benefit of fairness} more than others. For example, in attributes with a bimodal distribution, say, a mixture of people with no career breaks at all, and long career breaks (e.g., maternity etc.), people with mid-sized career breaks may get re-assigned, and could benefit or lose out depending on which side of the line they fall. Consider another example of a data pre-processing method for fair clustering; the fairlet clustering method~\cite{chierichetti2017fair}, in a gender-balanced dataset, would create fairlets as pairs, each pair comprising one from each gender (assuming binary genders for narrative simplicity only). Data objects that do not have an object of the other gender in its vicinity would stand to lose out due to being paired with a far-off object with which it bears shallow resemblance. As from the above two cases and their comparative evaluation, the cost of the fairness adjustments are unlikely to be {\it random} and would be borne asymmetrically across dataset objects. Higher volatility, and thus higher benefits or detriments, would likely be placed on objects that deviate much from the implicit data pattern assumptions made within the clustering formulations. While such qualitative differences of fairness shortfalls would be hard to be done away with, a characterization of the fairness shortfall, through quantitative metrics or exemplars, would be necessary to inspire confidence that fair clustering formulations do not exacerbate secondary biases while alleviating major ones. %This, as observed earlier, would be fairly critical within public sector scenarios. 

\section{Application Space Information}

The clusters in our job shortlisting scenario, we assumed, would be manually assigned one of three decisions, eventually leading to one of two decisions, shortlist or reject. Once this process is complete, we would obviously only care about whether there is representational parity on sensitive attributes over the {\it shortlisted} set (being just two sets, this would implicitly be equivalent to ensuring the same for the {\it rejected} set as well). In other words, the upstream clustering algorithm that tried to enforce representational parity in {\it each} of the several hundred clusters it generated, was, simply put, addressing a needlessly constrained problem. While it is impossible for the clustering algorithm to foresee the human decisions that would be assigned to each cluster, fair clustering formulations could be re-designed to provide interactive fairness guidance. For example, as soon as a cluster is chosen for the {\it shortlist} decision, the clustering could be re-run on the residual dataset with a different fairness target, that which seeks fairness among the clusters {\it conditional on the choice(s) already made} (this may be seen as similar in spirit the {\it alternative clustering} task~\cite{bae2006coala} at the high-level). Another handling of this would be for a one-shot clustering to produce, along with clusters, dependencies among clusters indicating that certain cluster pairs be assigned the same decision. This would be expected in cases of clusters that deviate from fairness in {\it different directions}, so this dependency constraint across them would help offset them. Such dependencies could also be envisioned as being one of {\it must-link} and {\it cannot-link} inspired by related literature on semi-supervised clustering~\cite{basu2002semi}. In cases with multiple (discrete/continuous) decision choices in the spectrum of positive to negative decisions, representational parity or other fairness considerations may be higher in certain parts of the decision space than others. For example, we may want to ensure that the set of failed candidates in a course not be very homogeneous on gender or race, whereas these may be more relaxed at the higher grades. In other words, 90\% of first graders being of a particular ethnicity may be more tolerable than an outcome where 90\% of fails coming from the same ethnic background. In short, information on clustering usage would go a long way in providing computational leeway in the pursuit of the chosen fairness targets for the clustering method. 

\section{Concluding Notes}

The above discussion was intended towards unravelling the diverse and inter-disciplinary possibilities in extending the scholarly frontier in fair clustering. We hope that researchers with interests in fair clustering would take note of such myriad research frontiers and diversify fair clustering research, an important task for data-driven decision making for the future. 

\bibliographystyle{named}
\bibliography{ijcai20}

\end{document}